\documentclass[manuscript]{aastex631}

\usepackage{graphicx}
\usepackage{subfigure} 
\usepackage{amsmath}
\usepackage{bm}
\usepackage{float}
\usepackage{placeins}
\usepackage[percent]{overpic}
\PassOptionsToPackage{left,mathlines}{lineno} 
\usepackage{lineno} 

\newif\ifshowrevisions
\showrevisionsfalse     

\begin{document}
\title{Parallel and perpendicular diffusion of energetic particles in the near-Sun solar wind observed by Parker Solar Probe}


\author[0009-0008-2509-9330]{Nibuna S. M. Subashchandar}
\affiliation{Department of Space Science, University of Alabama in Huntsville, Huntsville, AL, USA}

\author[0000-0002-4299-0490]{Lingling Zhao}
\affiliation{Department of Space Science, University of Alabama in Huntsville, Huntsville, AL, USA}
\affiliation{Center for Space Plasma and Aeronomic Research (CSPAR), University of Alabama in Huntsville, Huntsville, AL, USA}

\author[0000-0002-2923-0489]{Andreas Shalchi}
\affiliation{Department of Physics and Astronomy, University of Manitoba, Winnipeg, Manitoba R3T 2N2, Canada}

\author[0000-0002-4642-6192]{Gary Zank}
\affiliation{Department of Space Science, University of Alabama in Huntsville, Huntsville, AL, USA}
\affiliation{Center for Space Plasma and Aeronomic Research (CSPAR), University of Alabama in Huntsville, Huntsville, AL, USA}

\author[0000-0001-9199-2890]{Jakobus Le Roux}
\affiliation{Department of Space Science, University of Alabama in Huntsville, Huntsville, AL, USA}
\affiliation{Center for Space Plasma and Aeronomic Research (CSPAR), University of Alabama in Huntsville, Huntsville, AL, USA}

\author[0000-0003-3556-6568]{Hui Li}
\affiliation{Los Alamos National Laboratory, Los Alamos, NM 87545, USA}

\author[0000-0002-1541-6397]{Xingyu Zhu}
\affiliation{Center for Space Plasma and Aeronomic Research (CSPAR), University of Alabama in Huntsville, Huntsville, AL, USA}

\author[0000-0001-6286-2106]{Ashok Silwal}
\affiliation{Department of Space Science, University of Alabama in Huntsville, Huntsville, AL, USA}

\author[0000-0001-9581-3167]{Juan G Alonso Guzman}
\affiliation{Department of Space Science, University of Alabama in Huntsville, Huntsville, AL, USA}

\begin{abstract}
We investigate energetic particle diffusion in the inner heliosphere ($\sim$0.06--0.3 AU) explored by Parker Solar Probe (PSP). Parallel ($\kappa_\parallel$) and perpendicular ($\kappa_\perp$) diffusion coefficients are calculated using second-order quasi-linear theory (SOQLT) and unified nonlinear transport (UNLT) theory, respectively. PSP's in-situ measurements of magnetic turbulence spectra, including sub-Alfvénic solar wind, are decomposed into parallel and perpendicular wavenumber spectra via a composite two-component turbulence model. These spectra are then used to compute $\kappa_\parallel$ and $\kappa_\perp$ across energies ranging from sub-GeV to GeV. Our results reveal a strong energy and radial distance dependence in $\kappa_\parallel$. While $\kappa_\perp$ remains much smaller, it can rise accordingly in regions with relatively high turbulence levels $\delta B/B_0$. 
To validate our results, we estimate $\kappa_\parallel$ using upstream time-intensity profile of a solar energetic particle event observed by the PSP and compared it with theoretical values from different diffusion models. Our results suggest that the SOQLT-calculated parallel diffusion generally shows better agreement with SEP intensity–derived estimates than the classic QLT model. This indicates that the SOQLT framework, which incorporates resonance broadening and nonlinear corrections and does not require the introduction of an ad hoc pitch-angle cutoff, may provide a more physically motivated description of energetic particle diffusion near the Sun.

\end{abstract}

\section{Introduction} \label{sec:intro}
The transport of energetic charged particles in the heliosphere is fundamentally governed by their interactions with turbulent electromagnetic fields embedded in the solar wind. These turbulent fluctuations cause particles to scatter both parallel and perpendicular to the mean magnetic field, a process characterized by key transport parameters such as pitch-angle diffusion coefficients and scattering mean free paths. Accurate modeling of these diffusion processes is essential for understanding the propagation of solar energetic particles (SEPs), anomalous cosmic rays (ACRs), and galactic cosmic rays (GCRs) \citep{Giacalone2023}. A detailed understanding of the geometry and spectral properties of heliospheric turbulence, especially in the near-Sun environment, is critical for characterizing its influence on particle diffusion. The Parker Solar Probe (PSP) mission has provided unprecedented high-resolution in-situ measurements of magnetic field fluctuations close to the Sun, including in sub-Alfvénic regions where the solar wind speed is less than the local Alfvén speed \citep{kasper2021parker, zank2022turbulence, Zhao2022aApJ}. In the super-Alfvénic solar wind, Taylor’s hypothesis is typically employed to convert time-series measurements into spatial domain, under the assumption that the fluctuation velocity is much smaller than the solar wind speed. This allows the observed spacecraft-frame frequency spectra to be interpreted as wavenumber spectra, which are then used as direct inputs into particle diffusion models. Such approaches have been widely applied across the heliosphere, including in early PSP orbits \citep[e.g.,][]{Chen2024}. Despite recent observational advances, particle diffusion in the sub-Alfvénic region remains poorly understood, with perpendicular diffusion in the near-Sun environment notably absent from previous studies.

On the theoretical side, the quasi-linear theory (QLT) has long served as the standard framework for calculating particle diffusion coefficients \citep{Giacalone1999}. QLT treats particles as undergoing small perturbations around unperturbed Larmor orbits and assumes a delta-function resonance condition to describe wave–particle interactions. While analytically tractable, QLT exhibits a critical deficiency known as the $90^\circ$-scattering problem: as the pitch-angle cosine $\mu$ approaches zero, QLT requires resonance with waves of infinite wavenumber, which is unphysical, especially in steep turbulence spectra where power at small scales rapidly diminishes \citep{Qin2009,Qin2014,Shalchi2009}. To mitigate this singularity, studies often introduce a lower cutoff in pitch-angle cosine (e.g., $\mu_{\text{min}} = 0.05$) to enable approximate calculations \citep[e.g.,][]{Giacalone2020}. However, this workaround introduces model-dependent uncertainties, as the results are sensitive to the arbitrary choice of $\mu_{\text{min}}$ \citep[e.g.,][]{Li2022}. To address these limitations, second order quasi-linear theory (SOQLT) was developed as an extension of QLT that accounts for perturbations in particle orbits due to turbulent fields \citep{Shalchi2005}. By incorporating orbit fluctuations, SOQLT broadens the resonance condition, allowing particles to interact with a continuous range of wavenumbers rather than a singular resonance. This refinement makes SOQLT particularly suitable for application in regimes where the turbulence spectrum steepens at small scales such as the dissipation range, and where QLT's narrow resonance fails to capture realistic pitch-angle scattering. In addition, SOQLT resolves the non-integrable singularity at $\mu = 0$, improving the physical consistency of the theory and yielding better agreement with numerical simulations \citep{Qin2009, Qin2014}.

Beyond parallel diffusion, understanding perpendicular (cross-field) transport is also crucial, particularly in structured and anisotropic turbulence. The unified nonlinear transport (UNLT) theory offers a self-consistent extension of the Non-Linear Guiding Center (NLGC) \citep{Matthaeus2003} approach for modeling perpendicular diffusion \citep{Bieber2004,Shalchi2010}. UNLT correctly predicts zero cross-field diffusion in the absence of transverse fluctuations that is consistent with test-particle simulations, while QLT incorrectly predicts ballistic transport in all cases \citep{Qin2014, zhao2019generation, Shalchi2020}. Together, SOQLT and UNLT provide a more physically grounded framework for modeling particle diffusion in turbulent plasmas.
In this study, we apply SOQLT and UNLT to estimate the parallel and perpendicular diffusion coefficients of energetic particles in the inner heliosphere, using turbulence spectra derived from PSP magnetic field observations. Prior work has shown that the nature of solar wind turbulence varies significantly with heliocentric distance and solar wind parameters. In particular, PSP observations at radial distance close to the Sun indicate a substantially lower 2D-to-slab turbulence ratio compared to near-Earth conditions, suggesting a fundamental shift in turbulence geometry in the inner heliosphere \citep{bandyopadhyay2021geometry, zhao2022inertial}. Such geometric changes can significantly modify particle diffusion in the inner heliosphere. Moreover, in the sub-Alfvénic region, the conversion of spacecraft-frame frequency spectra to wavenumber spectra becomes nontrivial, as Taylor’s hypothesis may no longer be valid. To address these challenges, we use a two-component composite turbulence model to decompose the observed magnetic turbulence spectra in the sub-Alfvénic and moderate Alfvénic solar wind into parallel and perpendicular wavenumber spectra. These spectra are then incorporated into the SOQLT and UNLT frameworks to evaluate the energy and radial dependence of both parallel and perpendicular diffusion coefficients in the near-Sun environment. Finally, we validate our results by comparing the theoretically calculated $\kappa_\parallel$ with an empirical estimate of $\kappa_\parallel$ derived from time-intensity profile of an observed SEP event \citep{Giacalone2023, Lang2024}. This integrated observational–theoretical approach advances our understanding of turbulence-driven particle scattering in the inner heliosphere.

The paper is organized as follows. Section 2 reviews magnetic turbulence spectra observed by PSP within 0.3 AU. Section 3 presents results for both parallel and perpendicular diffusion coefficients for energetic particles ranging from from 500 keV to 1 GeV. Section 4 verifies the SOQLT-based $\kappa_\parallel$ predictions using the time-intensity profile of a SEP event observed by PSP. Section 5 provides a summary and discussion.


\section{Turbulence spectra close to the Sun}

\begin{figure}[htbp]
\centering
\includegraphics[width=0.7\textwidth]{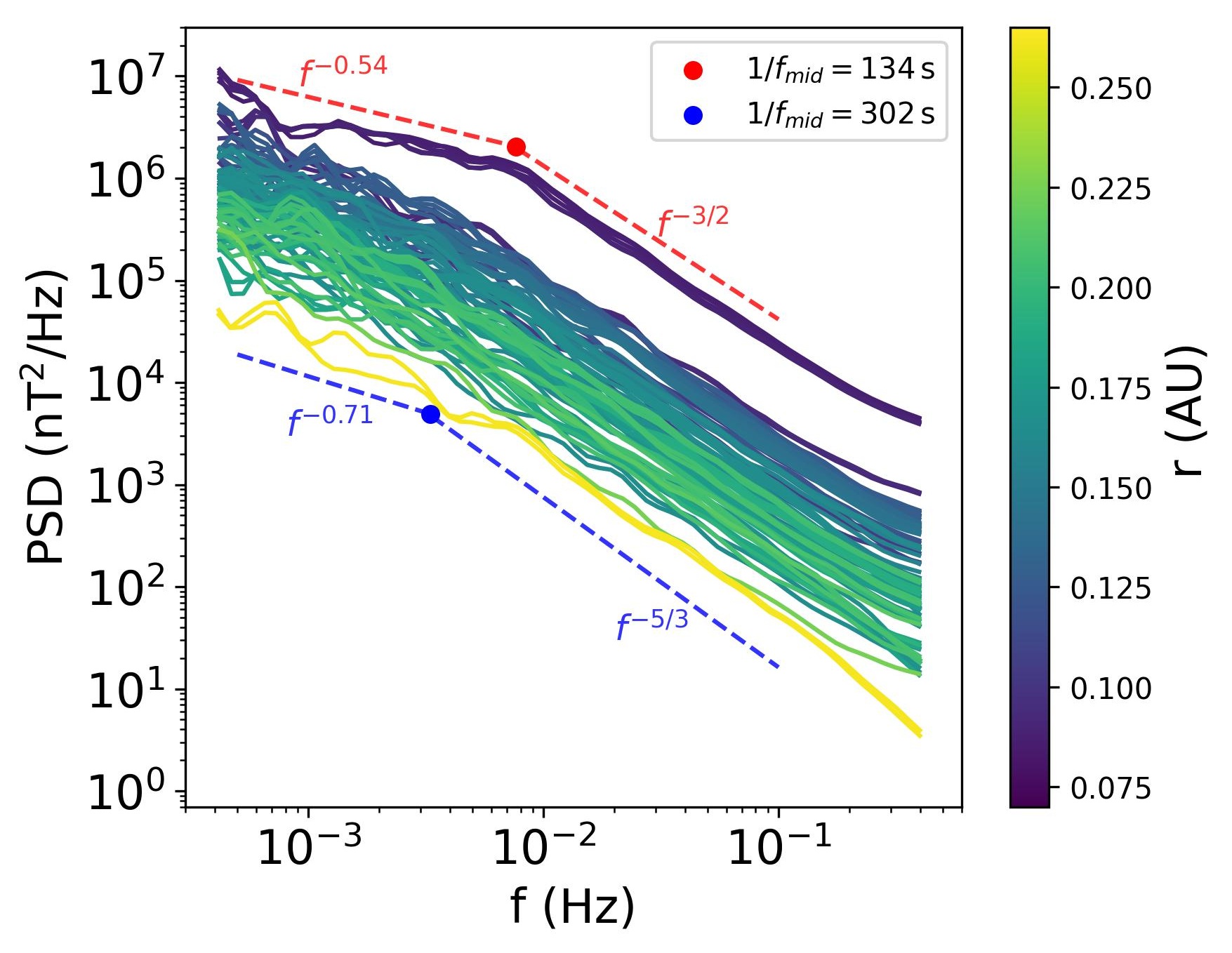}%
\caption{Power spectral density (PSD) of the transverse magnetic field fluctuations from PSP orbits E1-E19 observations with quasi-parallel ($\theta_{BV} < 15^{\circ}$) sampling. Each PSD is calculated with a 5-hour interval and color-coded by the radial distance $r$ from the Sun. Inertial range Kolmogorov-like $f^{-5/3}$ and Iroshnikov-Kraichnan $f^{-3/2}$ spectra are shown for reference. $f_{mid}$ represents the frequency separating the energy containing range and the inertial range and is determined as the statistical midpoint containing 50\% of the total fluctuation power.}
\label{fig:combined_plots}
\end{figure}
To first illustrate the spectral features of solar wind turbulence in the inner heliosphere, we compute the power spectral density (PSD) of magnetic field fluctuations using the averaged 1s cadence magnetic field data from PSP/FIELDS \citep{Bale2016SSRv}. These PSDs are calculated over nonoverlapping 5-hour intervals using wavelet analysis. To mitigate edge effects at low frequencies in wavelet analysis, all PSD calculations are restricted to regions within the cone of influence \citep{Huang2024}. 
Figure \ref{fig:combined_plots} presents the PSD of the transverse magnetic field fluctuations within 0.3 AU from PSP E1-E19 encounters, where the sampling angle $\theta_{BV}$ is less than 15 degrees, indicating that slab-like turbulence dominates \citep{Bieber1996}. $\theta_{BV}$ is the angle between the mean magnetic field and the bulk flow direction. The transverse fluctuations are determined with respect to the global mean magnetic field calculated over each 5-hour interval. 
As shown in the figure, the fluctuation power represented by the PSD systematically decreases with increasing distance from the Sun. The mid-frequency $f_{mid}$ that separates the inertial scale and energy containing range is defined as the statistical midpoint containing 50\% of the total fluctuation power \citep{Huang2024}. $f_{mid}$ shifts toward lower frequencies with increasing distance, indicating that the correlation length increases with radial distance. For two representative cases, $f_{mid}$ is about 1/134 Hz at about 0.07 AU (top spectrum), and about 1/302 Hz at about 0.3 AU (bottom spectrum). According to the Taylor hypothesis, this corresponds to a length scale of $8\times10^4$ km at about 0.07 AU, and a length scale of $1.3\times10^5$ km at about 0.3 AU. Within the inertial range, the spectral index transitions from approximately the Iroshnikov–Kraichnan scaling of $-3/2$ near the Sun toward the Kolmogorov scaling of $-5/3$ at larger distances. The steepening of the inertial range spectral index in parallel sampling ($\theta_{BV}\le 15^\circ$) indicates the parallel fluctuations gradually dissipate with distance, which may be due to the decay process or the weakening of nonlinear interactions in the magnetic field. For the energy containing range, the spectral index becomes steeper with distance and is slightly flatter than the -1 typically observed near the Earth \citep{Davis2023, Huang2024} for the 5-hour interval considered here.

In contrast, intervals dominated by perpendicular sampling (e.g., $\theta_{BV} \ge 75^\circ$), which are rare and not shown in this paper, do not exhibit a clear transition between the energy-containing range and the inertial range. This is probably due to insufficient statistical sampling in the perpendicular direction. Additionally, the spectral index in these intervals appears to be nearly independent of radial distance and remains close to the Kolmogorov-like scaling of $f^{-5/3}$. As the solar wind expands, the Parker spiral becomes more stretched in the azimuthal direction. Consequently, measurements at larger distances are expected to capture more intervals dominated by 2D turbulence (i.e., with perpendicular wavevector only), allowing for a better study of the 2D turbulence correlation length and its energy range spectral characteristics. We will not consider these very few perpendicular sampling intervals (no more than 10) in this paper.
The distinct features of magnetic fluctuation spectra observed near the Sun, in contrast to those at 1 AU, are expected to affect particle diffusion across different energy ranges depending on the underlying resonance conditions.

We use SOQLT to calculate the parallel diffusion coefficient $\kappa_\parallel$. Within this framework, the second-order pitch-angle Fokker–Planck coefficient \(D_{\mu \mu }^{( 2)}\) (see Appendix for details) is given by:

\begin{equation}
    D_{\mu \mu }^{( 2)} = \frac{4\pi \Omega^{2}\left( 1-\mu ^{2}\right)}{B_{0}^{2}} \int _{0}^{\infty } dk_{\parallel } \, g^{slab}( k_{\parallel }) \int _{0}^{\infty } dt\, \cos( \Omega t) \, e^{ik_{\parallel } v\mu t} e^{- k_{\parallel }^{2} \sigma _{z}^{2}( t) /2}
     \label{eq:dmumu}
\end{equation}
where \(v\) is the particle speed, \(\Omega\) is its gyro-frequency, \(\mu\) is the pitch-angle cosine, \( g^{slab}( k_{\parallel }) \) is the wavenumber spectrum of slab turbulence with parallel wavevector only, \(B_{0}\) is the mean magnetic field, 
and $\sigma_{z}$ is the width of the broadened resonance function, which acts as a correction to QLT. 

We note that the magnetic field data in Figure \ref{fig:combined_plots} were downsampled to a 1-second cadence, which imposes a Nyquist frequency of $\sim$0.5 Hz. Fluctuations above this frequency cannot be resolved. In a QLT framework with a $\delta$-function resonance, this would imply that low-energy particles at $\mu\approx0$ could not resonate, since their required high-wavenumber fluctuations may lie above the 0.5 Hz cutoff. This would artificially truncate $D_{\mu\mu}$ at $\mu\approx 0$. By contrast, the SOQLT formalism does not rely on a strict $\delta$-function resonance. Instead, it incorporates resonance broadening due to finite $\delta B/B_0$ and wave decorrelation, replacing the $\delta$-function with a finite-width kernel $\sigma_{z}$ in $\mu$ and $k$ \citep[e.g.,][]{Shalchi2005}. Scattering power is therefore distributed over a band of modes rather than concentrated at a single resonant mode, leaving $D_{\mu\mu}^{(2)}$ finite near $\mu\approx0$ even when the exact resonant wavenumber lies above the Nyquist limit.
In practice, the residual cadence effect is confined to a narrow neighborhood of $\mu\approx0$ and diminishes at higher energies, where particles resonate with lower-$k$ fluctuations. Consequently, the SOQLT-based $\kappa_{\parallel}$ remains robust across the energy range considered. The detailed derivation and its physical basis are summarized in the Appendix.

The parallel diffusion coefficient \(\kappa_{\parallel}\) is then obtained by integrating the pitch-angle diffusion coefficient over pitch angle $\mu$:
\begin{equation}
    \kappa_{\parallel } = \frac{v^{2}}{8} \int _{-1}^{1}\frac{\left( 1-\mu ^{2}\right)^{2}}{D_{\mu \mu }^{( 2)}} \, d\mu .
     \label{eq:parallelk}
\end{equation}

The perpendicular diffusion coefficient $\kappa_\perp$ is then calculated based on the Unified Non-Linear Transport theory (UNLT) model \citep{Shalchi2010}:

\begin{equation}
    \kappa _{\perp }  = \frac{\pi }{3}  \frac{a^{2} v^{2}}{B_{0}^{2}}  \int_{0}^{\infty} dk_{\perp }  \frac{g^{2D}(k_{\perp})}{( 4/3)  \kappa _{\perp } k_{\perp }^{2}  + v/\lambda _{\parallel }},
    \label{eq:UNLT}
\end{equation}
where the parallel mean free path (\(\lambda _{\parallel }\)) is determined from \(\kappa_{\parallel}\) (Equation \eqref{eq:parallelk}) using the relation \( \lambda _{\parallel } ={3} \kappa _{\parallel } /{v}\) and \( g^{2D}( k_{\perp }) \) is the wavenumber spectrum of 2D turbulence with perpendicular wavevector only. $a^2$ here serves as a correction factor to match theory with simulations. In this work, we adopt the conventional value \(a^2 = 1/3\) following \cite{Shalchi2013}.  For completeness, we note that more advanced formulations of perpendicular diffusion, specifically the Field Line Particle Decorrelation (FLPD) theory, developed by \cite{shalchi2021perpendicular}, eliminate the need for the free parameter \(a^2\), but is considerably more complex to calculate compared to UNLT model.

We decompose the observed inertial–range spectra in Figure~\ref{fig:combined_plots} with the two-component (i.e., slab and 2D components) turbulence spectral model \citep{Bieber1996, zank2022turbulence}:
\begin{equation}
    f\,\mathrm{P}_{\rm SUM}(f)
   = 2\,C_{\rm S}
     \Bigl(\frac{2\pi f}{V_{\rm sw}\cos\psi}\Bigr)^{1-q^{s}}
   + 2\,C_{\rm 2D}
     \Bigl(\frac{2\pi f}{V_{\rm sw}\sin\psi}\Bigr)^{1-q^{2D}},
     \label{bieber_inertial_composition}
\end{equation}
where \(C_{\mathrm S}\) and \(C_{\mathrm{2D}}\) are fluctuation power of the slab and 2D components, respectively. \(V_{\!\rm sw}\) is the solar wind speed in the spacecraft frame, \(\psi\) is the acute angle between the mean magnetic field and the bulk flow direction, \( q^{s, 2D} \) is the spectral index of slab and 2D turbulence, respectively. $f$ is the measured frequency and \( f\ \text{P}_{\rm SUM}(f) =   f\ P_{\mathrm S}(f)\ +\ f\  P_{\mathrm{2D}}(f) \) is the measured total fluctuation energy. We assume the inertial range turbulence spectral index $-q^s = 1.5$ for the slab component and $-q^{2D} = 5/3$ for the 2D component. Based on Equation \eqref{bieber_inertial_composition}, each measured inertial range frequency spectrum in Figure \ref{fig:combined_plots} is fitted using a least-squares method to determine the coefficients $C_s$ and $C_{2D}$. This allows decomposing the frequency spectrum observed in each 5-h interval into the sum of the slab and 2D turbulence spectra. To calculate the particle diffusion coefficients, their corresponding wavenumber spectra are required as input. The mapping from frequency to wavenumber spectrum depends on the Alfvénic regime. For 2D turbulence, which is non-propagating and passively convected by the solar wind \citep{zhao2023observations, zhao2025non}, the conversion is straightforward via the Taylor hypothesis, i.e., $k_\perp = \frac{2\pi f}{V_{\rm sw} \sin\psi}$, and thus $\mathrm{P}_{\rm 2D}(k_\perp) = 2\,C_{\rm 2D}\,k_\perp^{-q^{\rm 2D}}$. For the slab component, the classical Taylor hypothesis,
$k_\parallel = \frac{2\pi f}{V_{\rm sw} \cos\psi}$ is valid only for strongly super-Alfvénic flows, i.e., \(M_A \equiv \dfrac{V_{\rm sw}}{V_A} \gg 1,\ \text{where } V_A \text{ is the Alfvén speed}\). For moderately Alfvénic or sub-Alfvénic intervals that are commonly observed during later PSP encounters (e.g., E8–E14), we adopt a modified Taylor hypothesis that accounts for plasma-frame wave propagation \citep{Zhao2024ApJ}. This approach converts the slab frequency spectrum into its wavenumber counterpart using the following relation, $k_\parallel = \frac{2\pi f}{V_{\rm sw} \cos\psi + V_A}$ \citep{zank2022turbulence}. The resulting slab wavenumber spectrum is expressed as
$\mathrm{P}_{\rm S}(k_\parallel) = 2\,C_{\rm S}\,k_\parallel^{-q^{\rm S}}$. In practice, we evaluate \(M_A\) for each interval. Intervals with $M_A > 5$ are classified as strongly super-Alfvénic, a regime where the classical Taylor formulation is used. For intervals where $M_A \le5$, we apply the modified Taylor relations described above \citep[see also][]{zank2022turbulence, Zhao2022_E10, zhu2025radial}. A practical consideration for the modified Taylor hypothesis near perihelion is the role of spacecraft motion. In this study, we used SWEAP proton moments reported in the spacecraft frame. Consequently, the effective advection speed entering the Doppler shift relation is taken directly from these measurements, and no additional subtraction of the spacecraft velocity is required. This is consistent with the recommendations of \citet{Klein2015}, who emphasized correcting for spacecraft motion only when plasma velocities are expressed in a Sun-centered frame. Furthermore, we verify that near perihelion the effective advection speed in our intervals is $\sim 200$–$300~\mathrm{km\,s^{-1}}$, well above the local Alfvén speed. This confirms that the validity conditions for the modified Taylor hypothesis are satisfied.

\begin{figure}[t]
\centering
\includegraphics[width=0.6\textwidth]{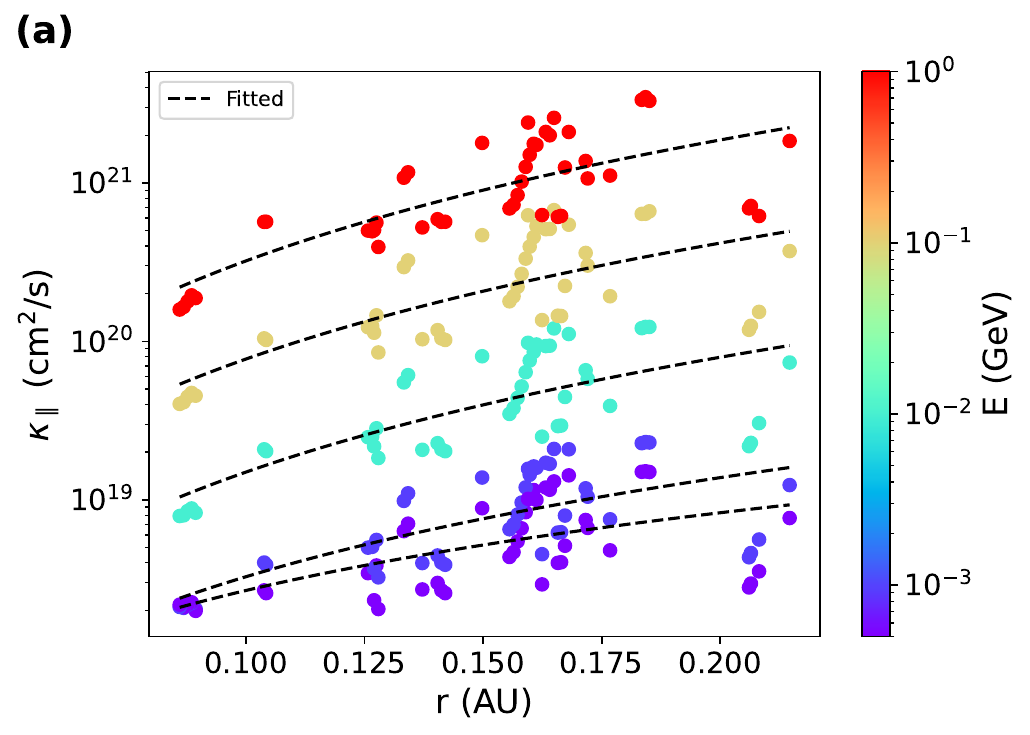}\\
\includegraphics[width=0.6\textwidth]{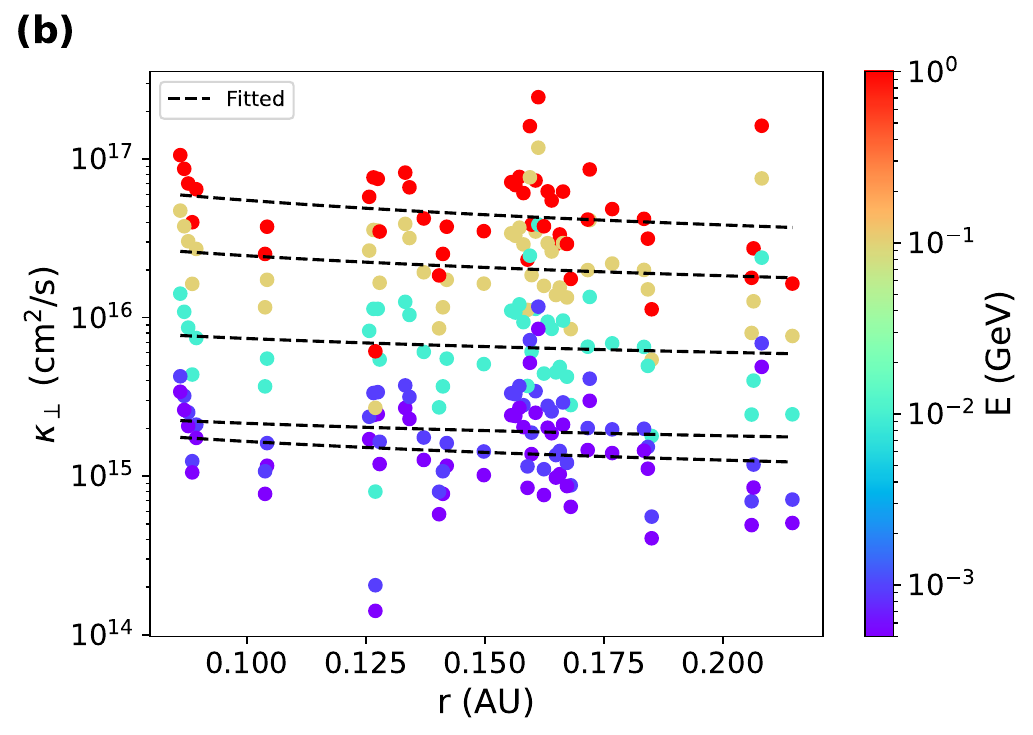}
\caption{Radial evolution of the parallel and perpendicular diffusion coefficients, $\kappa_\parallel$ (SOQLT) and $\kappa_\perp$ (UNLT), derived from the observed turbulence spectra. (a) $\kappa_\parallel$ as a function of radial distance from the Sun (10–60 $R_\odot$), color-coded by particle kinetic energy (500 keV to 1 GeV).
(b) $\kappa_\perp$ displayed similarly.
In both panels, black dashed lines represent power-law fits obtained via linear least-squares regression of the scatters.}
\label{fig:kappa_valid_bieber_radial_evolution}
\end{figure}
\section{Particle parallel and perpendicular diffusion coefficients}
Due to the predominantly parallel sampling geometry in this near-Sun regime (within 0.3 AU), the relative contribution of 2D turbulence is significantly smaller compared to that of slab turbulence. The decomposed slab component inertial range frequency spectra are approximately one to two orders of magnitude (i.e., 10–100 times) higher in power than the corresponding 2D turbulence spectrum (not shown here). This suggests a dominant slab contribution to the total turbulence energy in the analyzed intervals in Figure \ref{fig:combined_plots}.
The obtained slab wavenumber spectrum $P_s(k_\parallel)$ is then substituted into Equations \eqref{eq:dmumu} and \eqref{eq:parallelk}, yielding the parallel diffusion coefficient $\kappa_\parallel$. Similarly, the obtained 2D wavenumber spectra $P_{2D}(k_\perp)$ is fed into Equation \eqref{eq:UNLT} to determine the perpendicular diffusion coefficient $\kappa_\perp$.
As shown in Figure \ref{fig:combined_plots}, the turbulence spectrum varies with distance, and the resulting $\kappa_\parallel$ and $\kappa_\perp$ also vary with radial distance in addition to the particle kinetic energy $E$.

Figures \ref{fig:kappa_valid_bieber_radial_evolution}(a) and (b) show the dependence of the parallel and perpendicular diffusion coefficients on radial distance and particle kinetic energy, respectively. Panel (a) presents the SOQLT-based parallel diffusion coefficient $\kappa_{\parallel}$, while panel (b) shows the UNLT-based perpendicular diffusion coefficient
$\kappa_{\perp}$. The analysis focuses on the near-Sun region, from approximately 10 to 60 solar radii, and considers particle kinetic energies ranging from 500 keV to 1 GeV, which primarily interact with inertial range turbulence. The scatters, derived from PSP magnetic field measurements and the decomposed wavenumber spectra, are color-coded by particle energy $E$. Black dashed lines represent power-law fits obtained via linear least-squares regression of the
scatters. The radial dependence of $\kappa_\parallel$ varies with particle energy. For example, for 500 keV protons, it follows approximately $r^{1.1}$, while for 1 GeV protons, it increases more steeply as $r^{1.5}$. In contrast, $\kappa_\perp$ exhibits a much weaker dependence on particle energy. Across the five energy ranges analyzed, the radial scaling of $\kappa_\perp$ remains nearly constant, approximately following $r^{-0.13}$. Regarding the weak or absent radial dependence of $\kappa_{\perp}$ in Figure~\ref{fig:kappa_valid_bieber_radial_evolution}(b), we attribute this to a coupled effect that arises naturally from the radial evolution of the fluctuation energy and the parallel diffusion. We compute $\kappa_{\perp}$ using UNLT as in Equation~(\ref{eq:UNLT}). The numerator of Equation~(\ref{eq:UNLT}) contains the 2D turbulence spectrum $g^{2\mathrm{D}}(k_{\perp},r)$, which represents the 2D fluctuation energy and typically decreases with heliocentric distance approximately as $1/r$ \citep[e.g.,][]{zhao2017cosmic}. Meanwhile, the parallel mean free path in the denominator increases roughly as $\lambda_{\parallel}\propto r$. To leading order, these opposing scalings cancel within the integrand of Equation~(\ref{eq:UNLT}), leaving $\kappa_{\perp}$ only weakly dependent on $r$, as reflected in Figure~\ref{fig:kappa_valid_bieber_radial_evolution}(b). This relatively weak radial dependence of perpendicular diffusion is consistent with earlier studies based on Nonlinear Guiding Center theory \citep[e.g.,][]{Pei2010,zhao2017cosmic}. The relatively small magnitude of $\kappa_\perp$ (approximately $10^{-4}$-$10^{-3}$ of $\kappa_\parallel$) can be attributed to the fact that the observed frequency spectra in Figure \ref{fig:combined_plots} predominantly reflect parallel sampling, where $\theta_{BV} \le 15^\circ$. This sampling configuration, which is statistically most common within 0.3 AU, leads to a significantly reduced 2D turbulence component when applying the spectral decomposition method developed by \cite{Bieber1996}.
However, we emphasize that in certain regions such as near the heliospheric current sheet (HCS) or in corotating interaction regions (CIRs), more oblique or quasi-perpendicular sampling is likely. Under these circumstances, the 2D turbulence fraction can increase, potentially enhancing the perpendicular diffusion coefficient $\kappa_\perp$. In fact, studies of CIRs show that perpendicular diffusion may become dominant in modulating cosmic-ray intensity in such regions \citep{AlonsoGuzman2025}. Despite these localized enhancements, particle transport remains overwhelmingly dominated by parallel diffusion. Our results reveal that, in the radial range of 10 to 60 solar radii under primarily parallel-sampling conditions, $\kappa_\parallel$ increases by nearly an order of magnitude, while $\kappa_\perp$ varies by less than $\sim40\%$. This pronounced diffusion-tensor anisotropy results in a smaller ratio of $\kappa_\perp/\kappa_\parallel$ at closer heliocentric distances.

\begin{figure}[t]
\centering
\includegraphics[width=0.6\textwidth]{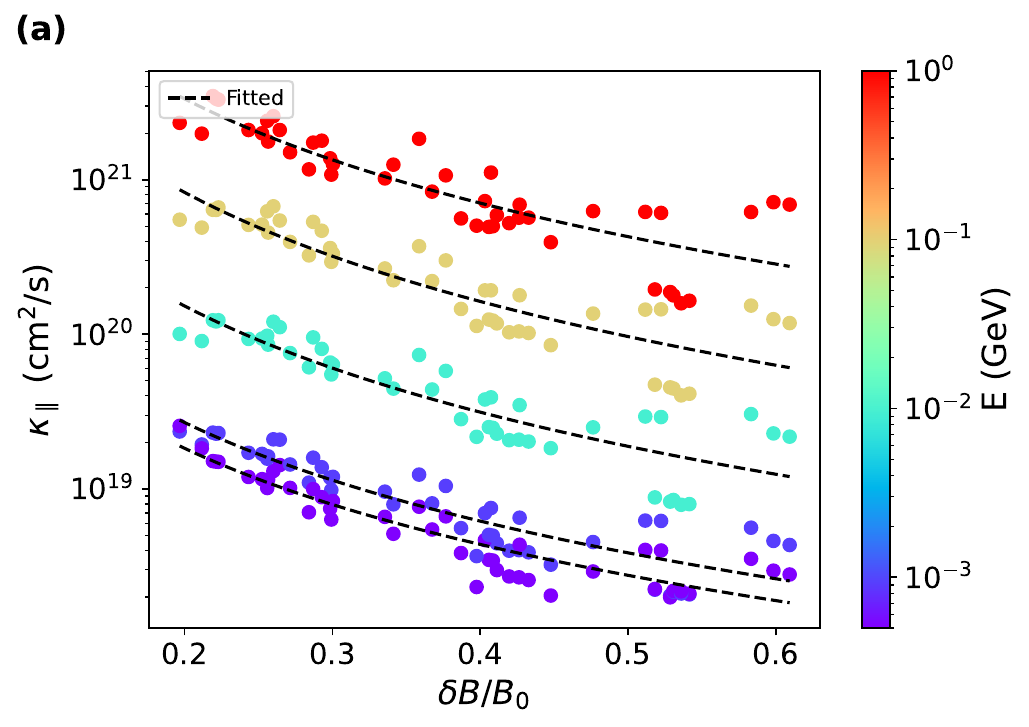}\\
\includegraphics[width=0.6\textwidth]{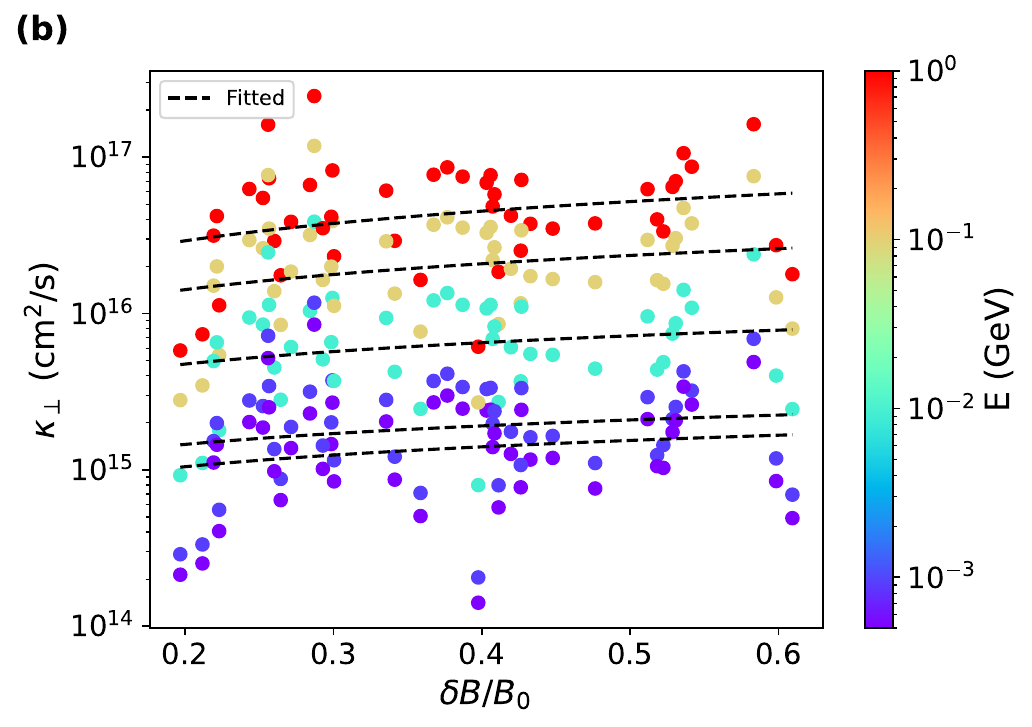}
\caption{Radial evolution of the diffusion coefficients 
$\kappa_\parallel$ (SOQLT) and $\kappa_\perp$ (UNLT) as functions of the turbulence amplitude $\delta B/B_0$. Panels (a) and (b) show 
$\kappa_\parallel$ and $\kappa_\perp$, respectively, with data points color-coded by particle energy (500 keV to 1 GeV). Black dashed lines indicate power-law trends obtained by linear least-squares fitting.}
\label{fig:kappa_valid_bieber_delta_b_b0_evolution}
\end{figure} 

Since turbulence spectra quantify the power in magnetic fluctuations, diffusion coefficients naturally scale with fluctuation amplitude. At 1 AU, several empirical expressions based on Kolmogorov-like inertial spectra and QLT model capture how $\kappa_\parallel$ and $\kappa_\perp$ scale with normalized fluctuation amplitude $\delta B/B_0$ \citep{Giacalone1999, zhao2017cosmic, zhao2018influence}. It is found that the parallel diffusion coefficient $\kappa_\parallel$ decreases with increasing $\delta B/B_0$ \citep[e.g.,][]{zhao2014modulation}, because stronger fluctuations enhance scattering and reduce particle propagation along field lines. Conversely, the perpendicular diffusion coefficient $\kappa_\perp$ increases with $\delta B/B_0$, since enhanced turbulence promotes field-line wandering and cross-field transport. Despite this, $\kappa_\perp$ generally remains much smaller than $\kappa_\parallel$, though its relative contribution grows in more turbulent regimes.
To quantify this, we integrate the PSD in Figure \ref{fig:combined_plots} to obtain $\delta B^2$, and calculate the relative fluctuation amplitude $\delta B/B_0$, where $B_0$ is the mean magnetic field averaged over each 5‑hour interval. 

Figures \ref{fig:kappa_valid_bieber_delta_b_b0_evolution}(a) and (b) show how the SOQLT parallel diffusion coefficient 
$\kappa_\parallel$ and the UNLT perpendicular diffusion coefficient $\kappa_\perp$ vary with the turbulence amplitude 
$\delta B/B_0$. It's clear that parallel diffusion decreases with the increasing fluctuation amplitude and perpendicular diffusion slightly increases with $\delta B/B_0$. In panel (a), the decay exponent for $\kappa_\parallel$ remains approximately constant at $-2.13$ across all considered proton energies, indicating a scaling relation of $\kappa_\parallel \sim (\delta B/B_0)^{-2.13}$ within the fluctuation amplitude range of roughly 0.2 to 0.6 observed by PSP. In panel (b), the UNLT perpendicular diffusion coefficient $\kappa_\perp$ exhibit the opposite trend, which increases with turbulence amplitude. At a fixed $\delta B/B_0$, the perpendicular diffusion $\kappa_\perp$ increase with particle energy roughly as $E^{0.5}$. The scaling exponent for $\kappa_\perp$ with $\delta B/B_0$ flattens from $\sim$0.63 at 1 GeV to $\sim$0.40 at 500 keV, indicating a weaker dependence of perpendicular diffusion on turbulence amplitude at lower energies.
From Figure \ref{fig:kappa_valid_bieber_delta_b_b0_evolution}, both diffusion coefficients are sensitive to turbulence amplitude, the decreasing trend of $\kappa_\parallel$ with $\delta B/B_0$
is considerably stronger than the increasing trend observed for $\kappa_\perp$. In essence, increasing turbulence significantly suppresses parallel transport, while only modestly enhancing perpendicular diffusion, suggesting that parallel diffusion remains the dominant transport mechanism in the inner heliosphere, especially given that $\kappa_\parallel$ typically exceeds $\kappa_\perp$ by about three orders of magnitude.


\begin{figure}[htbp]
\centering
\includegraphics[width=0.5\textwidth]
{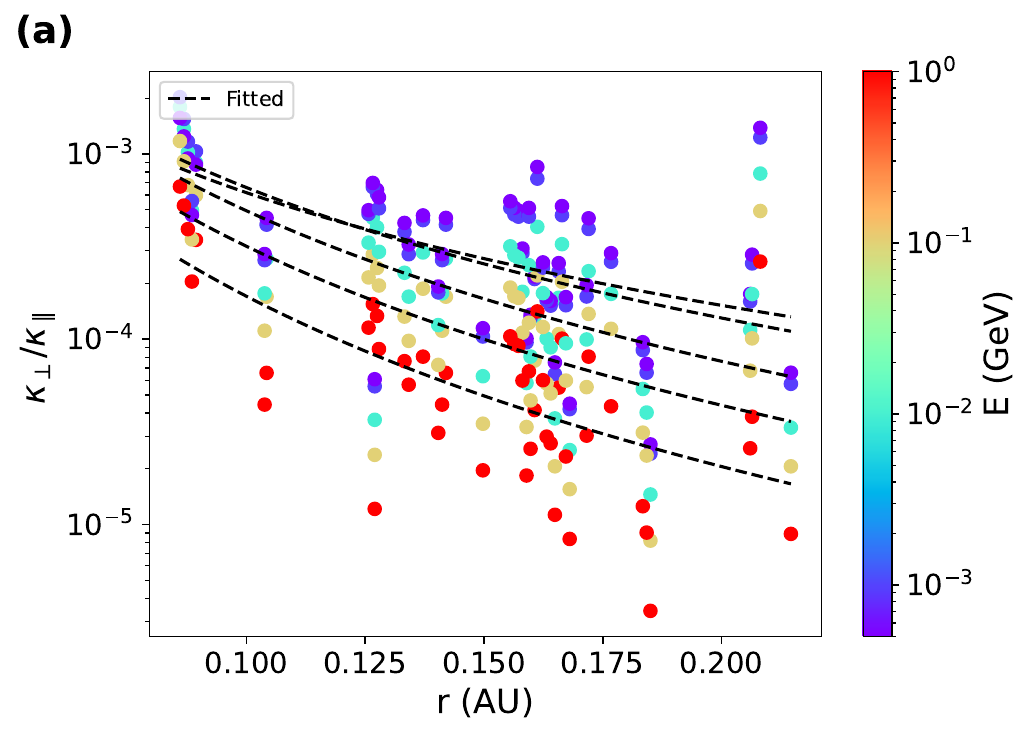}%
\includegraphics[width=0.5\textwidth]
{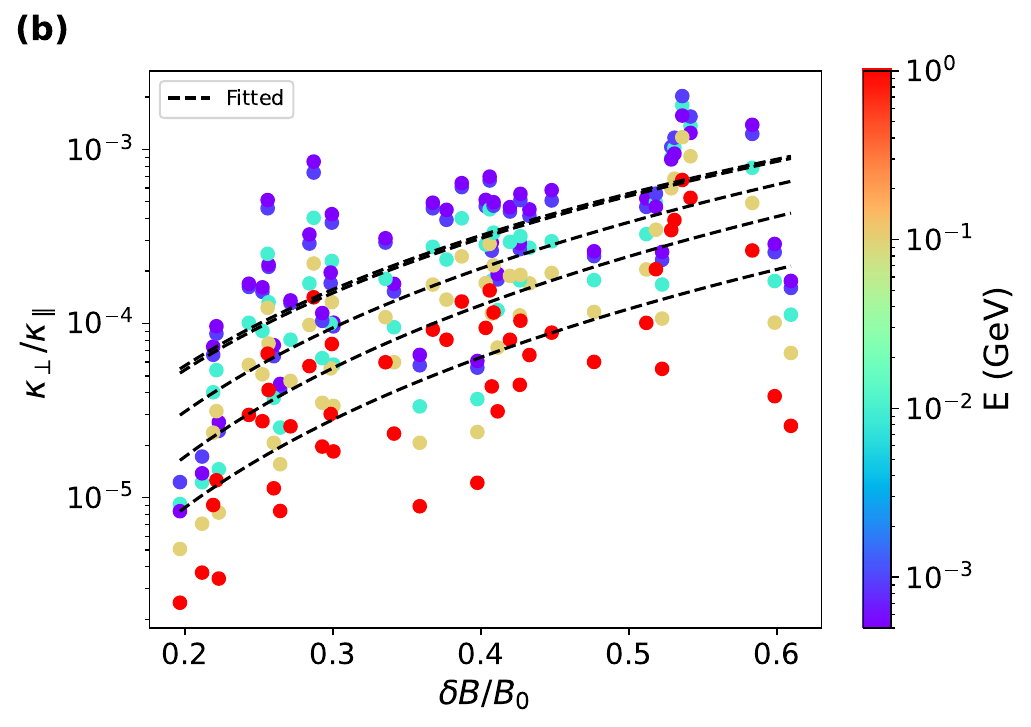}
\caption{Dependence of the diffusion coefficient ratio $\kappa_\perp/\kappa_\parallel$ on (a) heliocentric distance $r$, and (b) the relative fluctuation amplitude $\delta B/B_0$. Scatter points are derived from PSP E1–E19 observations, and the black dashed curves show power-law fits for five distinct particle energy ranges.}  
      
\label{fig:kappa_ratio_valid_bieber_delta_b_b0_evolution}
\end{figure}

In Figure \ref{fig:kappa_ratio_valid_bieber_delta_b_b0_evolution}, we plot the ratio $\kappa_\perp/\kappa_\parallel $ against the radial distance in panel (a) and relative fluctuation amplitude in panel (b). Black dashed curves indicate power-law fits across five energy ranges.
As shown in panel (b), $\kappa_\perp/\kappa_\parallel $ increases steadily with fluctuation amplitude $\delta B/B_0$. For fluctuation amplitude rising from $\sim$0.2 to $\sim$0.6, $\kappa_\perp/\kappa_\parallel $ spans roughly an order of magnitude across all energy ranges. At a fixed fluctuation amplitude $\delta B/B_0$, the ratio $\kappa_\perp/\kappa_\parallel $ is largest at lowest energy 500 keV and smallest at 1 GeV. This behavior suggests stronger suppression of $\kappa_\parallel$ by turbulence at low energies compared to higher energies.
In contrast, panel (a) shows the ratio 
$\kappa_\perp/\kappa_\parallel $ as a function of heliocentric distance $r$. The ratio clearly decreases with increasing radial distance. For the lowest energy range 500 keV, the radial decline is comparably flatter, due to the relatively slower increase of $\kappa_\parallel$
at lower energies, as illustrated in Figure 2(a). Overall, the ratio $\kappa_\perp/\kappa_\parallel$ scales with radial distance approximately as $r^{-1.2}$ for 500 keV protons and steepens to about \(r^{-1.6}\) 
for higher energy protons. Figure \ref{fig:kappa_ratio_valid_bieber_delta_b_b0_evolution} (a) and (b) together demonstrate that local turbulence and radial expansion act in opposite directions on the diffusion-tensor anisotropy, i.e., strong turbulence increases the ratio $\kappa_\perp/\kappa_\parallel$, whereas heliocentric expansion decreases $\kappa_\perp/\kappa_\parallel$. 
The relatively small $\kappa_\perp/\kappa_\parallel$ ratio shown in Figure \ref{fig:kappa_ratio_valid_bieber_delta_b_b0_evolution} may result from the limited sampling of 2D turbulence during PSP's perihelion crossing. Since 2D fluctuations have wavevectors largely perpendicular to the mean magnetic field, they are less well resolved by PSP measurements, leading to an underestimation of the 2D turbulence amplitude. Because perpendicular diffusion $\kappa_\perp$ depends strongly on this amplitude, undersampling these modes produces smaller inferred $\kappa_\perp$ and, consequently, a reduced $\kappa_\perp/\kappa_\parallel$ ratio. In certain intervals, such as near the HCS or CIRs, the inferred 2D turbulence fraction is higher, allowing $\kappa_\perp$ to increase correspondingly.

\section{Validation of the parallel diffusion}

\begin{figure}[t]
\centering
\includegraphics[width=0.8\textwidth, keepaspectratio]{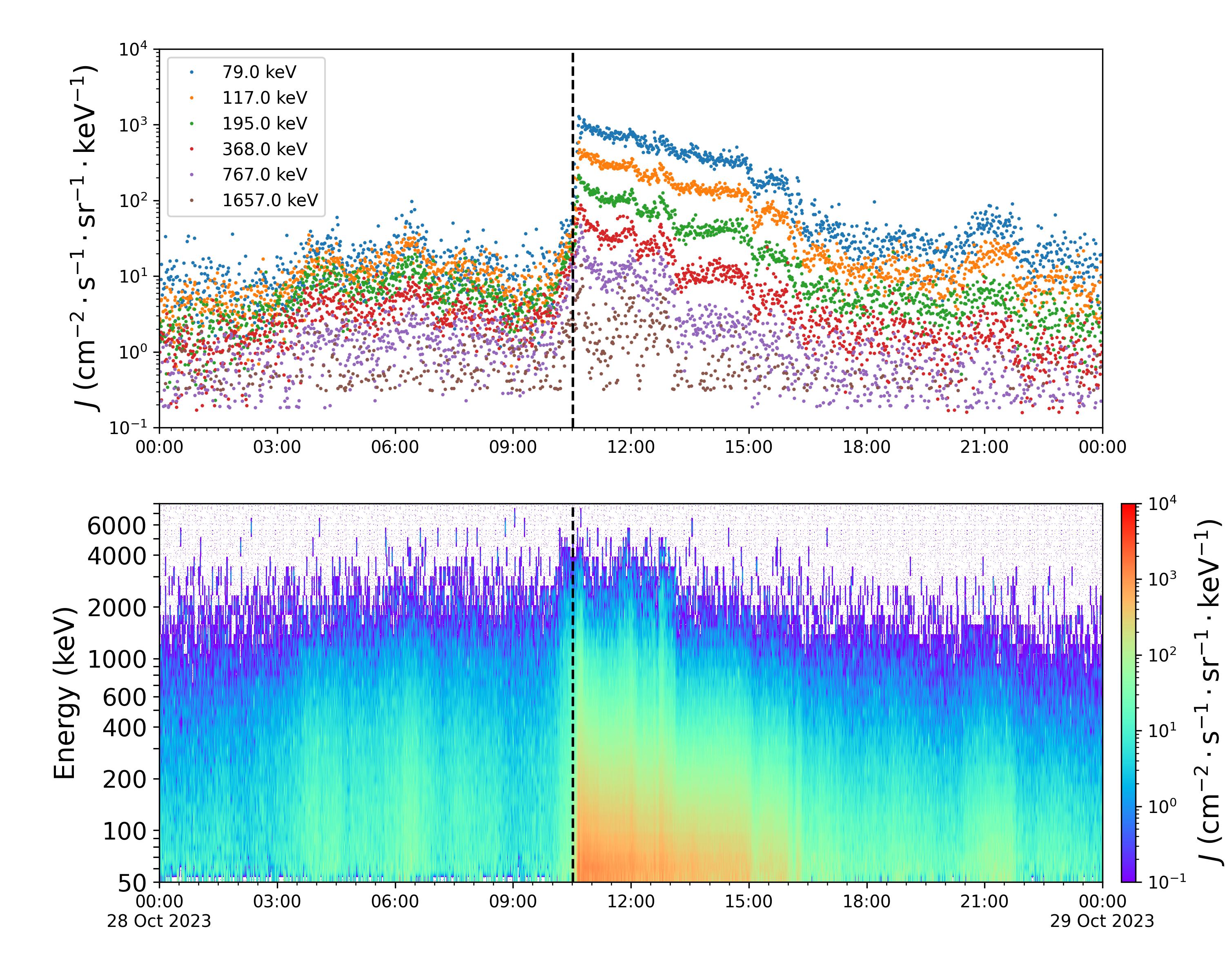} 
\caption{Overview of the solar energetic particle (SEP) event observed by PSP on October 28, 2023. The Top panel shows the differential intensity of energetic protons across multiple energy channels. The bottom panel shows a spectrogram of differential intensity for all analyzed protons, with energy on the vertical axis and intensity indicated by color bar. The vertical dashed line marks the arrival of the interplanetary shock.}
    \label{fig:Feb_16_EP_event}
\end{figure}

As a benchmark and for comparison with the parallel diffusion predicted by QLT, we estimate the parallel diffusion coefficient from the observed time-intensity profiles of energetic particles. Figure \ref{fig:Feb_16_EP_event} shows an energetic proton event observed by PSP/IS$\odot$IS \citep{mccomas2016integrated} on October 28, 2023 near an interplanetary shock at a distance of about 0.7 AU. The top panel in Figure \ref{fig:Feb_16_EP_event} shows the differential intensity of energetic protons ranging from 79 keV to 1657 keV, and the bottom panel shows the spectrogram of the differential flux, with proton energy on the y-axis. The vertical dashed line indicates the shock arrival time at around 10:31:34 UT. The shock is a fast forward shock driven by an ICME. The density compression ratio is about 3.1, the shock obliquity $\theta_{Bn}$ is about 66 degrees, and the shock speed is about 556 km/s in the observer (spacecraft) frame. The shock normal $\hat{\mathbf{n}}$ in the RTN coordinate system is $\hat{\mathbf{n}} = (0.82, 0.37, -0.44)$. We used a mixed-coplanarity method \citep{AbrahamShrauner1976} to estimate the shock normal direction and combined it with a mass flux algorithm to calculate the shock velocity \citep{zhao2021turbulence}, which gives results consistent with those reported by \citet{Kruparova2025} within uncertainties.

As shown in the top panel, the flux across all energy channels rises exponentially immediately upstream of the shock, indicating diffusive particle transport toward the spacecraft. We fit the intensity profile using $I(t) = I_0 \exp\left(\frac{t}{\Delta t}\right)$, where $\Delta t$ is the exponential rise time of the upstream particle fluxes and is obtained via least-squares fitting. The upstream interval is taken as 15 minutes ahead of the shock, following \citet{Giacalone2023}. This choice balances statistical robustness by including more data points while remaining within the particle intensity exponential rise phase. We find that the 15-minute interval (10:15--10:30) provides an overall better exponential fit across all five energy channels, with goodness-of-fit $R$ values typically above 0.8, except for the lowest energy channel (75 keV), indicating strong agreement between the fitted intensity rise curves and the observed intensity profiles. Under diffusive assumptions, the diffusion length $L$, determined via Parker-spiral magnetic connectivity to the shock, relates to $\Delta t$ through $L^2 \simeq 2 \kappa_\parallel \Delta t$. Alternatively, following an empirical model \citep{giacalone2012energetic, Giacalone2023}, the diffusion coefficient along the shock propagation $\kappa_{rr}$ can be estimated using the approximation $\kappa_{rr} = W_1 V_{\text{sh}} \Delta t$, where \( W_1 \) is the upstream flow velocity along the shock normal direction in the shock rest frame and \( V_{\text{sh}} \) is the shock propagation speed. For the event studied here, \( V_{\text{sh}}=556 \,km/s \) and \( W_1 = 286 \,km/s \). The above empirical approximation yields the shock-normal diffusion coefficient $\kappa_{rr}$ rather than the field-aligned coefficient $\kappa_{\parallel}$. In this analysis, we apply a correction factor of $1/\cos^2(\theta_{Bn})$ to relate $\kappa_{rr}$ to $\kappa_{\parallel}$ and assume that $\kappa_{\perp}$ is small and neglected, i.e., $\kappa_{\parallel} = \kappa_{rr}/\cos^2(\theta_{Bn})$ \citep[e.g.,][]{zhao2017cosmic}.

\begin{figure}[t]
\centering
\includegraphics[width=0.42\textwidth]{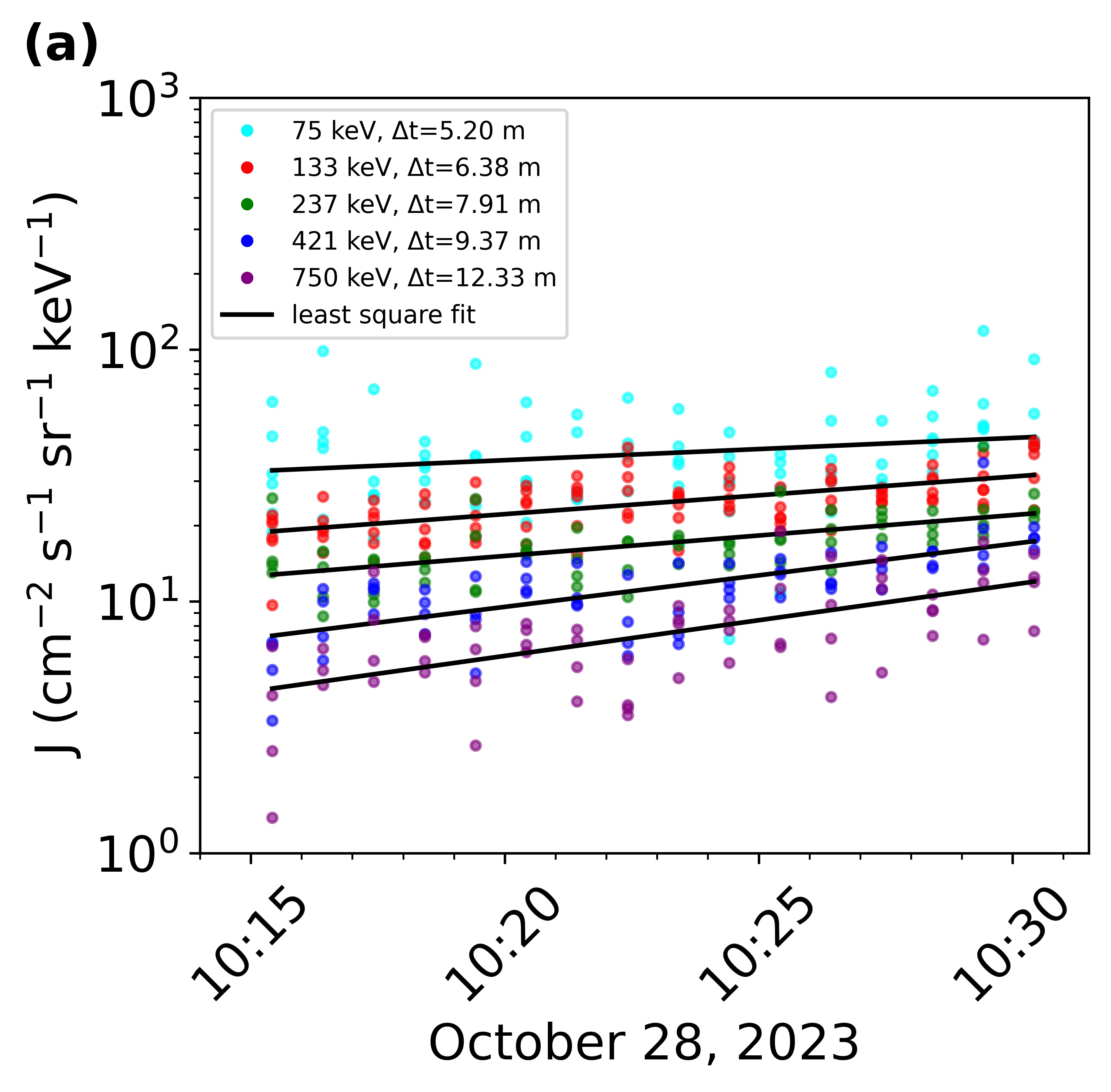}%
\includegraphics[width=0.51\textwidth]{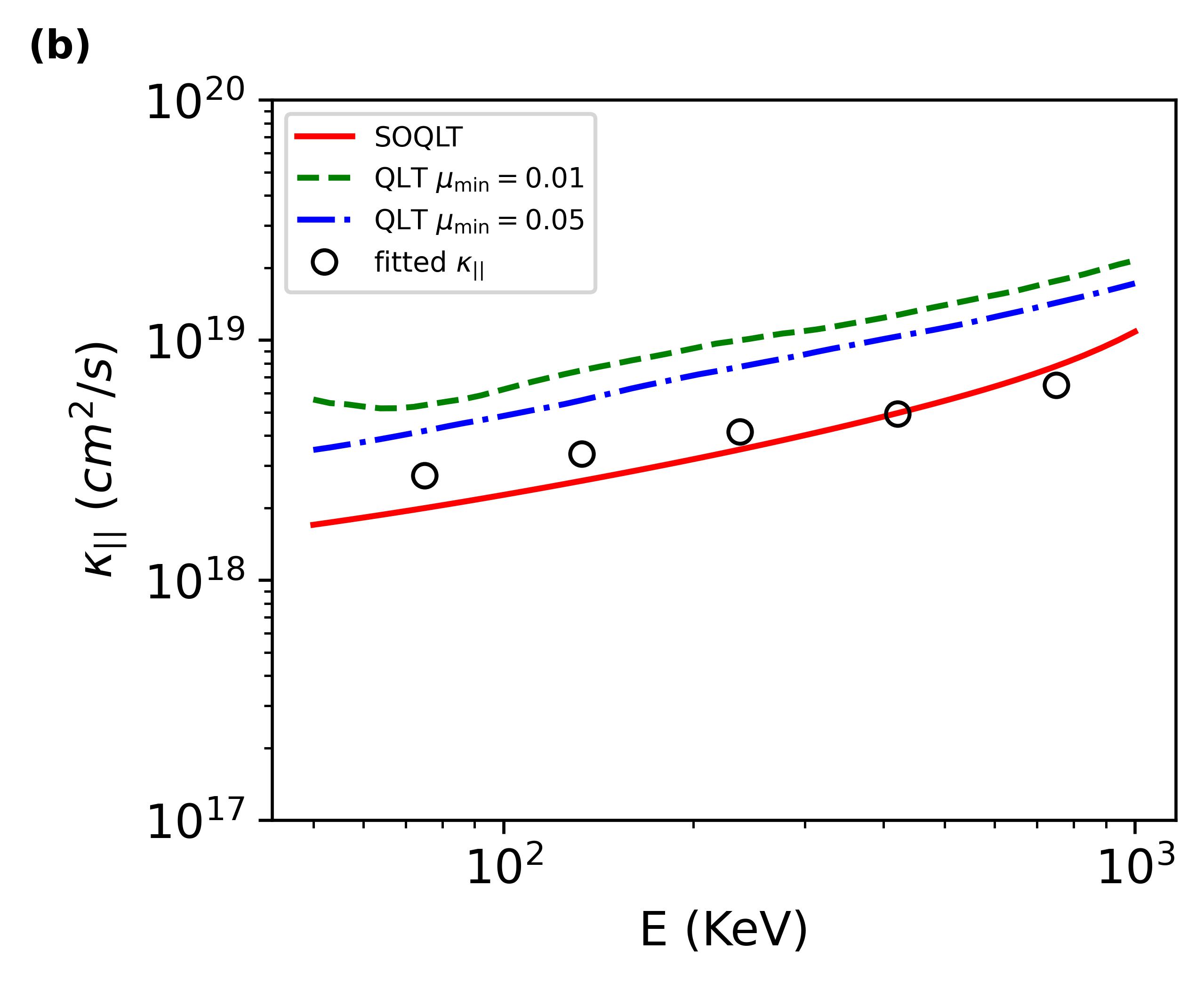}
\caption{\small Panel (a) shows the PSP/IS$\odot$IS EPI-Lo observed ion differential intensity (dots) over a 15-minute interval upstream of the shock on 2023 October 28. The black lines represent least squares fits to particle time-intensity profile to obtain the exponential rise time $\Delta t$ in minutes for each energy bin. The right panel shows the comparison between parallel diffusion coefficient calculated from QLT with assumptions (green and blue curves) and SOQLT (red curve) and particle time-intensity profile (red open circles).}
\label{fig:shock}
\end{figure}

Figure \ref{fig:shock}(a) shows exponential fits applied to the upstream 15-minute interval (approximately 10:15–10:30 UT) across five energy channels (75–750 keV), yielding an energy dependent rise time $\Delta t$, which is 5.2 minutes at 75 keV and 12 minutes at 750 keV.  These $\Delta t$ values, listed for each energy channel in the figure, reflect diffusive particle transport and serve as inputs for estimating the parallel diffusion coefficient $\kappa_\parallel$. In panel (b), we plot fitted $\kappa_\parallel$ (black open circles) values derived from $\Delta t$, shock speed $V_{sh}$, and upstream flow normal component $W_1$ in the shock frame, using the empirical approximation $\kappa_{rr}$. We then compare these fitted values to theoretical predictions, i.e., standard QLT with two minimum pitch-angle cosine thresholds ($\mu_{min}=0.05$ and 0.01) and SOQLT (red curves). We construct the turbulence spectrum from magnetic-field measurements over the same upstream interval used to estimate the exponential rise time $\Delta t$. Accordingly, both QLT and SOQLT calculations utilize the upstream 15-minute interval immediately preceding the shock ($\sim$\,10{:}15--10{:}30~UT), which is identical to the time window used in Figure \ref{fig:shock}(a) to derive the energy-dependent exponential rise time $\Delta t$.

As discussed in \cite{Li2022}, a minimum pitch-angle cosine threshold is required for QLT to link pitch-angle and spatial diffusion coefficients, due to the rapid falloff of the observed magnetic power spectrum at high frequencies caused by turbulence dissipation. This effect influences particle scattering near \( 90^\circ \) pitch angles. For this analysis, minimum pitch-angle cosine values $\mu_{min}$ of 0.05 and 0.01 are applied and compared with the results of SOQLT in Figure \ref{fig:shock}(b). As shown in panel (b), SOQLT aligns closely with the fitted $\kappa_\parallel$, especially at higher energies. In contrast, both QLT results, using $\mu_{min}=0.05$ (blue dashed-dotted curve) and $\mu_{min}=0.01$ (green dashed curve), exhibit noticeable deviations from the fitted $\kappa_\parallel$. Overall, SOQLT shows much better agreement with the fitted $\kappa_\parallel$. We also quantified the relative difference between the fitted $\kappa_\parallel$ and the theoretical values $\kappa_\parallel^T$ from QLT (under two assumptions of $\mu_{\min}$) and SOQLT using
$F = \frac{|\kappa_\parallel^T - \kappa_\parallel^{\rm fit}|}{\kappa_\parallel^{\rm fit}}$.
The $F$ values for SOQLT are close to zero across all five studied energy channels: 0.26 for 75 keV, 0.23 for 133 keV, 0.17 for 237 keV, 0.01 for 421 keV, and 0.18 for 750 keV. In contrast, the $F$ values for QLT, with either choice of $\mu_{\min}$, are significantly larger, approaching or exceeding 1. In addition, SOQLT avoids the problematic pitch-angle cosine cutoff at $\mu = 0$ inherent in QLT, although a slight discrepancy remains at lower energies, likely due to uncertainties in determining $\Delta t$. The diffusion coefficients shown in Figure~\ref{fig:shock}(b) are event-localized values near the shock, derived from the upstream 15-minute interval immediately preceding the crossing. They are not intended to represent typical background solar-wind diffusion coefficients reported in most statistical studies. In the upstream foreshock, streaming SEPs can self-generate resonant Alfv\'enic turbulence \citep[e.g.,][]{trotta2023irregular, zhao2025theory}, enhancing $D_{\mu\mu}$ and thereby reducing $\lambda_{\parallel}$ and $\kappa_{\parallel}$ relative to ambient values \citep[e.g.,][]{Lee2005ApJS}.

The Palmer (1982) consensus \citep{palmer1982transport}, derived from 1\,AU observations of cosmic rays and solar energetic protons, provides a benchmark parallel mean free path ($\lambda_{\parallel}$) of $0.08$--$0.3$\,AU for particles with rigidities of $0.5$--$5000$\,MV. This consensus, however, is limited to near-Earth conditions, averaging over different solar-wind regimes and based on relatively sparse datasets from the 1970s--1980s. Near the Sun, turbulence is stronger, more anisotropic, and highly intermittent, and the magnetic-field geometry is more complex. Simply extrapolating Palmer’s 1\,AU values inward would therefore overestimate the mean free path, because the near-Sun and near-Earth regions correspond to fundamentally different turbulent regimes. In our PSP intervals, SOQLT predicts smaller $\kappa_{\parallel}$ than QLT and the Palmer consensus. This is a physical consequence: classic QLT employs a $\delta$-function resonance that predicts vanishing pitch-angle diffusion $D_{\mu\mu}$ at $\mu=0$ (the $90^{\circ}$ scattering problem), which artificially suppresses scattering through $90^{\circ}$ and inflates $\kappa_{\parallel}$ unless an ad hoc $\mu_{\min}$ cutoff is applied. SOQLT, by broadening the resonance, restores finite $90^{\circ}$ scattering and accelerates pitch-angle isotropization, leading to larger $D_{\mu\mu}$ and hence smaller $\kappa_{\parallel}$. Therefore, the smaller $\kappa_\parallel$ predicted by SOQLT relative to QLT and the Palmer Consensus reflects both the proper inclusion of $90^{\circ}$ scattering and the properties of inner heliospheric turbulence sampled by PSP, rather than a breakdown of the theory. As shown in Figure \ref{fig:shock}, the SOQLT-predicted $\kappa_{\parallel}$ is also closer to the estimates derived from energetic particle transport, highlighting the need to revisit the Palmer consensus for the near-Sun environment using careful inversion analyses of the observed solar energetic particle events.

\section{Conclusions}
We systematically study energetic particle diffusion in the inner heliosphere ($\le$0.3 AU) using Parker Solar Probe measurements, calculating the parallel diffusion ($\kappa_\parallel$) via SOQLT model and the perpendicular diffusion ($\kappa_\perp$) coefficients via UNLT model. By decomposing the in situ measured turbulence spectra into the wavenumber spectra of the slab and 2D turbulence, we derive the parallel and perpendicular diffusion coefficients for energetic particles ranging from sub-GeV to GeV. Their respective dependence on heliocentric distance, fluctuation amplitude, and particle energy are studied. The validity of our results is confirmed by comparison with the parallel diffusion coefficient obtained by fitting the time-intensity profile of the observed solar energetic particle event. The main findings are summarized as follows.

1. The parallel diffusion coefficient $\kappa_\parallel$ increases strongly with heliocentric distance, scaling approximately as $r^{1.1-1.5}$ depending on particle energy. The rapid increase of $\kappa_\parallel$ with distance may be due to the weakening of turbulence with distance. In contrast, perpendicular diffusion $\kappa_\perp$ remains nearly constant or even shows a slight decrease with increasing distance, likely due to the narrow heliocentric range studied and limited sampling of perpendicular fluctuations in the data. Nevertheless, radial distance plays a much more pronounced role in shaping $\kappa_\parallel$ than $\kappa_\perp$ in the inner heliosphere.

2. The parallel diffusion $\kappa_\parallel$ scales inversely with turbulence amplitude, approximately as $(\delta B/B_0)^{-2.13}$. This is consistent with the idea that stronger fluctuations enhance pitch-angle scattering, significantly reducing parallel transport efficiency. Perpendicular diffusion $\kappa_\perp$ increases with turbulence level, albeit less dramatically, following roughly $(\delta B/B_0)^{0.6}$. This moderate scaling reflects perpendicular diffusion benefits of more turbulent mixing, but does not escalate as dramatically as parallel suppression.

3. Perpendicular diffusion $\kappa_\perp$ remains a very small fraction of parallel diffusion in the region close to the Sun, largely due to PSP’s predominantly parallel sampling. Our results show that $\kappa_\perp$ can be three to four orders of magnitude smaller than $\kappa_\parallel$. However, for lower-energy particles at smaller heliocentric distances and in regions of elevated turbulent fluctuations, the relative contribution from $\kappa_\perp$ increases appreciably. We caution that the extremely small $\kappa_\perp/\kappa_\parallel$ observed is likely an observational bias due to the sampling effects. In more turbulent environments, such as in corotating interaction regions (CIRs) or near the heliospheric current sheet (HCS), $\kappa_\perp$ can be largely enhanced. Nevertheless, parallel diffusion remains the dominant transport mechanism in the near-Sun environment.

4. The parallel diffusion coefficient $\kappa_\parallel$ derived from fitting the upstream time–intensity rise during an observed SEP event agrees well with SOQLT predictions, while showing significant discrepancies with both standard QLT results. Unlike classical QLT, which requires an artificial pitch-angle cosine cutoff ($\mu_{min}>0$) to avoid the singularity at $\mu=0$, SOQLT incorporates resonance broadening and nonlinear corrections that naturally regulate behavior near $\mu=0$. This makes SOQLT the preferred method for accurately modeling parallel diffusion close to the Sun.

We present, for the first time, the evolution of both parallel and perpendicular diffusion in the near-Sun environment (including the sub-Alfvénic solar wind) and demonstrate their variation with radial distance, particle energy, and fluctuation level. We emphasize that the long-standing $90^\circ$ scattering problem in QLT limits its applicability, except perhaps at very high particle energies. Although one may introduce a free parameter, e.g., a lower cutoff $\mu_{min}$, to fit observed time–intensity profiles, the physical meaning and determination of $\mu_{min}$ are unclear and may vary from event to event. This flexibility can make QLT appear to behave similarly to SOQLT or even to the fitted $\kappa_\parallel$ in some energy ranges, but the introduction of $\mu_{min}$ is unphysical, as it effectively imposes an ad hoc lower bound on pitch-angle scattering. In this context, $\mu_{min}$ functions as a free fitting parameter rather than a quantity derived from first principles. In contrast, SOQLT naturally accounts for the $90^\circ$ scattering problem through resonance broadening, without requiring any adjustable parameter.

Finally, we note that extracting a proper $\kappa_\parallel$ from particle time–intensity profiles requires applying appropriate transport models \citep{Lang2024}. The empirical formula we used \citep{giacalone2012energetic} assumes that the upstream particle intensity profile is exponential, governed by the one-dimensional balance between convection and radial diffusion. This framework further assumes sufficiently strong scattering conditions, under which particle propagation can be approximated by spatial diffusion. In this regime, the exponential rise time $\Delta t$, determined from fitting the observed time–intensity profile, directly encodes the upstream diffusion coefficient. However, this approach neglects additional effects such as turbulence evolution upstream of the shock and deviations from isotropy in the particle distribution. We caution that a more rigorous determination of the diffusion coefficient requires either physically complete diffusion theories (e.g., SOQLT) based on the observed turbulence spectrum, or proper inversion of the particle transport model, accounting for all relevant transport effects \citep{Lang2024}.

In summary, decomposing the observed frequency spectra into parallel and perpendicular wavenumber spectra enables a direct modeling of energetic particle diffusion. 
By combining this spectral decomposition with advanced transport theories, i.e., SOQLT for 
$\kappa_\parallel$ and UNLT for $\kappa_\perp$, and using PSP’s in-situ turbulence measurements near the Sun, we implement more comprehensive diffusion modeling in the inner heliosphere. Extending our study to environments with enhanced turbulence and magnetic complexity, such as CIRs and the HCS, will be crucial as perpendicular diffusion can be significantly boosted by stronger turbulence and field-line meandering in these regions.

\vspace{1pt}

\section{acknowledgments}
We thank the NASA Parker Solar Probe team for the use of data. We acknowledge the partial support of the NSF award 2442628, NASA awards 80NSSC20K1783, 80NSSC23K0415, and 80NSSC24K1867, and a NASA Parker Solar Probe contract SV4-84017, NSF EPSCoR RII-Track-1 Cooperative Agreement OIA-2148653. A.S. acknowledges support by the Natural Sciences and Engineering Reseach Council (NSERC) of Canada. J.l.R. acknowledges support from NASA grant Nos. 80NSSC21K1319 and 80NSSC25K7750.
\vspace{15pt}

\appendix{}
\vspace{-15pt}
\section*{Theoretical overview of second order quasi-linear theory}
When charged particles propagate through a magnetized plasma, their trajectories are continuously perturbed by interactions with turbulent magnetic field fluctuations. Each interaction causes a small deflection in the particle’s direction of motion. Over time, the cumulative effect of these deflections leads to a diffusive spreading of the particle’s position in space \citep{Mario2019}. 
The transport of energetic particles along magnetic field lines is described by a two-dimensional Fokker–Planck equation in space and pitch-angle cosine, capturing the combined effects of parallel streaming and pitch-angle scattering \citep{Schlickeiser2002, Zank2014}.
\begin{align}
    \frac{\partial f}{\partial t} + v {\mu} \frac{\partial  f}{\partial z} &= \frac{\partial }{\partial \mu} \left( D_{\mu\mu} \frac{\partial  f}{\partial \mu} \right) \label{eq:pfeq}.
\end{align}
Here, $f(t,z,\mu)$ denotes the particle distribution function, $z$ represents the spatial coordinate parallel to the mean magnetic field $\bm{B}$, $v$ is the particle velocity, and $\mu=\bm{v}\cdot \bm{B}/(vB)$ is the cosine of the particle pitch angle. The term $D_{\mu\mu}$ corresponds to the pitch-angle Fokker–Planck coefficient (or pitch-angle diffusion coefficient), quantifying the diffusion of particles in pitch-angle space.
Each term in the equation defines a distinct characteristic timescale: the evolutionary timescale ($f/\partial_t f$), the spatial crossing timescale ($f/(|\partial_z f| v)$), and the pitch-angle scattering timescale ($1/D_{\mu\mu}$). The equation can be solved through successive approximation, employing the assumption that the evolutionary process occurs at the slowest rate, whereas pitch-angle scattering tends to isotropy on the fastest timescale. At zeroth order, Eq. \eqref{eq:pfeq} simplifies to
\begin{equation}
    0 \approx \frac{\partial }{\partial \mu} \left( D_{\mu\mu} \frac{\partial f}{\partial \mu} \right).
\end{equation}
Consequently, the anisotropy present in the particle distribution function $f$ must be minimal. At the lowest order, the distribution function is isotropic and independent of $\mu$. We therefore express the distribution function as a perturbative expansion:
\[
f(t,z,\mu) = f_0(t,z) + f_1(t,z,\mu) + f_2(t,z,\mu), \quad \text{with} \quad f_2 \ll f_1 \ll f_0.
\]
At first order, the governing equation becomes
\begin{equation}
    v\mu \frac{\partial f_0}{\partial z} = \frac{\partial }{\partial \mu} \left( D_{\mu\mu} \frac{\partial f_1}{\partial \mu} \right).
\end{equation}
Performing an integration over $\mu$ from $-1$ to $\mu$, and applying the boundary condition $D(\mu=\pm 1)=0$\footnote{Refer to Eq.\eqref{eq:Dotmu}: at $\mu=\pm 1$, we have $\dot{\mu}=0$, and thus Eq.\eqref{eq:Dmumu} implies $D(\mu=\pm 1)=0$.}, yields
\begin{equation}
    \frac{\partial f_1}{\partial \mu} = \frac{\mu^2-1}{2D_{\mu\mu}} v \frac{\partial f_0}{\partial z}. \label{eq:f1}
\end{equation}
The resulting expression relates $f_1$ explicitly to $f_0$. Subsequently, at second order—describing the slower evolution of the isotropic component $f_0$
\begin{equation}
    \frac{\partial f_0}{\partial t} + v\mu \frac{\partial f_1}{\partial z}
   = \frac{\partial }{\partial \mu} \left( D_{\mu\mu} \frac{\partial f_2}{\partial \mu} \right).
\end{equation}
We integrate over the entire range of $\mu$ from $-1$ to $+1$ to eliminate the term on the right-hand side, giving
\begin{align}
    \frac{\partial f_0}{\partial t} &= -\frac{v}{2} \frac{\partial}{\partial z}\int_{-1}^{+1} \mu f_1 d\mu \\
    &=  \frac{v}{2} \frac{\partial}{\partial z}\int_{-1}^{+1} \left( \frac{1}{2}\frac{\partial (1-\mu^2)}{\partial \mu} \right) f_1 d\mu.
\end{align}
Substituting Eq.~\eqref{eq:f1} and performing integration by parts leads to a standard diffusion equation:
\begin{equation}
    \frac{\partial f_0}{\partial t} = \frac{\partial }{\partial z}\left(\kappa_{\parallel}\frac{\partial f_0}{\partial z} \right).
\end{equation}
From this equation, we explicitly identify the spatial diffusion coefficient as
\begin{equation}
\kappa_{\parallel}  = \frac{v^2}{8} \int_{-1}^{+1} \frac{(1-\mu^2)^2}{D_{\mu\mu}} d\mu.
\label{eq:kappa}
\end{equation}
Once the pitch-angle diffusion coefficient $D_{\mu\mu}$ is specified, the spatial diffusion coefficient $\kappa_{\parallel}$ can be readily calculated. 

In the following, we rederive the expression for $D_{\mu\mu}$ within the framework of second-order quasilinear theory (SOQLT), which extends classical quasilinear theory by incorporating finite-amplitude effects and resonance broadening \citep{Shalchi2005,Shalchi2009}.
The pitch angle diffusion coefficient $D_{\mu \mu }$ can be calculated by the TGK (Taylor-Green-Kubo) formulation,
\begin{equation}
    D_{\mu \mu} = \int_{0}^{\infty} dt \ \langle \dot{\mu}(t) \ \dot{\mu}(0) \rangle \label{eq:Dmumu}.
\end{equation}
To calculate the pitch angle cosine
$\mu = \dfrac{v_{\parallel }}{v} $, we can apply the equation of motion for charged particles in a magnetic field,
\begin{equation}
    \frac{d\vec{p}}{dt} = q\left(  \frac{\vec{v}}{c} \times \vec{B} \right)\label{eq:motion},
\end{equation}
where we ignore the electric fields due to the high conductivity of plasma. Assuming that the magnetic field has the mean magnetic field $\bm B_0$ along the z axis and a turbulent component $\delta \bm B$ perpendicular to $\bm B_0$, Eq \eqref{eq:motion} can be written as 
\begin{gather}
    \dot{v}_{x} = \Omega \left( -v_{y} - v_{z}\frac{\delta B_{y}}{B_{0}} \right) ;\\
    \dot{v}_{y} = \Omega \left( v_{x} +  v_{z}\frac{\delta B_{x}}{B_{0}} \right) ;\\
    \dot{v}_{z} = \Omega \left( v_{x}\frac{\delta B_{y}}{B_{0}} - v_{y}\frac{\delta B_{x}}{B_{0}}\right), \label{eq:dvdz}
\end{gather}
where $\Omega=B_0q/(\gamma m)$ is the relativistic gyro frequency.
Thus the time evolution of the pitch angle cosine is 
\begin{equation}
    \dot \mu = \frac{\dot v_{\parallel}}{v} = \frac{\Omega}{v} \left[ v_{x} \frac{\delta B_{y}}{B_{0}} - v_{y} \frac{\delta B_{x}}{B_{0}} \right]. \label{eq:dotmu}
\end{equation}
By approximating $v_{x}$ and $v_{y}$ with the unperturbed particle velocities,
\begin{equation}
    v_{x} = v \sqrt{1 - \mu^{2}} \cos(\phi_{0} - \Omega t) \quad \text{and} \quad v_{y} = -v \sqrt{1 - \mu^{2}} \sin(\phi_{0} - \Omega t), \label{eq:vxvy}
\end{equation}
allows Equation \eqref{eq:dotmu} be simplified as
\begin{align}
    \dot{\mu} ( t) \ &=\ \frac{{\Omega}\sqrt{1-\mu ^{2}}}{B_{0}}[ \delta B_{y}(\vec{x} ,t) \ cos( \phi _{0} \ -\ {\Omega}t) \ +\ \ \delta B_{x}(\vec{x} ,t) \ sin( \phi _{0} \ -\ {\Omega}t)]. \label{eq:Dotmu}
\end{align}
Substituting into Eq \eqref{eq:Dmumu}, we obtain
\begin{align}
    D_{\mu \mu} = \frac{\Omega^{2} (1 - \mu^{2})}{B_{0}^{2}} \int_{0}^{\infty} dt \, \langle \delta B_{x}(t) \delta B_{x}^{*}(0) \rangle \big[ & \{ \cos(\phi_{0}) \cos(\Omega t) + \sin(\phi_{0}) \sin(\Omega t) \} \cos(\phi_{0}) \notag \\
    & + \{ \sin(\phi_{0}) \cos(\Omega t) - \cos(\phi_{0}) \sin(\Omega t) \} \sin(\phi_{0}) \big]  \label{eq:Dmumuphi}
\end{align}
for axisymmetric turbulence with vanishing magnetic helicity ($\langle \delta B_i \delta B_j \rangle =1/2 \langle \delta B^2 \delta_{ij}\rangle$). Here, $\mu=\mu(t=0)$ is treated as a constant. This holds under the assumption that the turbulence is sufficiently weak such that variation in $\mu$ remains small. However, this assumption breaks down over long timescales, as cumulative interactions can lead to significant changes in the particle's pitch angle.
We then average over the initial gyrophase $\phi_{0}$ by applying $\frac{1}{2}\int_0^{2\pi} d\phi$ to Eq \eqref{eq:Dmumuphi}, giving the simplified form
\begin{align}
    D_{\mu \mu } \ &=\ \frac{{\Omega}^{2}\left( 1-\mu ^{2}\right)}{B_{0}^{2}}\int _{0}^{\infty } dt\ cos( {\Omega}t) \ \langle \delta B_{x}( t) \delta B_{x}^{*}( 0) \rangle \label{eq:Dmumuf}
\end{align}
Equation (\ref{eq:Dmumuf}) relates the pitch angle diffusion coefficient to the correlation function of the turbulent magnetic field. On using the Fourier representation for $\delta B_x(\bm x)$, the correlation function can be written as  \citep{Shalchi2009,Zank2014}
\begin{equation}
    \langle \delta B_{i}(\vec{x} ,t) \delta B_{j}^{*}( 0,0) \rangle  = \int d^{3} k d^3 k'\ \langle \delta B_{i}(\vec{k} ,t) \delta B_{j}^{*}(\vec{k} ,0)  e^{i\vec{k} \cdot \bm x( t)}\rangle,
\end{equation}
where $\bm x$ is the particle's position. For slab turbulence, the turbulent fluctuations propagate along the mean magnetic field with $\bm k = k\bm e_z$. The exponential term can be written as $e^{i\bm k \cdot \bm x} = e^{ikz}$.
In QLT, the unperturbed particle position $z_{QLT}= v \mu t$ is used to replace the exact position. By approximating $\mu$ as a constant, $e^{ikz}$ can be factored out from the ensemble average.
However, in general, the integrand can be expressed as a product of the spectral tensor ($P$) and the characteristic (phase) function ($\Gamma$) by applying the Corrsin's independence hypothesis (\cite{Shlien1974}), 
\begin{equation}
    \langle \delta B_{i}(\vec{k}, t) \delta B_{j}^{*}(\bm {k'}, 0) e^{i\bm {k} \cdot \bm {x}(t)} \rangle \approx \langle \delta B_{i}(\bm {k}, t) \delta B_{j}^{*}(\bm {k'}, 0) \rangle \langle e^{ikz} \rangle = P_{ij}(\bm k,t)\delta(\bm k-\bm k')\Gamma(\bm {k}, t).
\end{equation}
Here, $P$ represents the distribution of kinetic energy in spectral space.
For the magnetostatic turbulence, the spectral tensor is independent of time, thus $P_{ij}(\bm {k} ,t)  = P_{ij}(\bm {k})$. For turbulence with slab geometry, we have $P_{ij}(\vec{k})  = g( k_{\parallel })\dfrac{\delta ( k_{\perp })}{k_{\perp }}$ and $g$ is power spectrum of slab turbulence.

 Using cylindrical coordinates to calculate the volume element of the wave vector as $d^3k=2\pi k_{\perp} dk_{\perp}dk_{\parallel}$, the correlation function becomes
\begin{align}
   \langle \delta B_{x}(t) \delta B_{x}^{*}(0) \rangle &= 4\pi \int_{0}^{\infty} dk_{\parallel} \, g(k_{\parallel}) \, \Gamma(k_{\parallel}, t),
   \label{eq:corr_fun}
\end{align}
with the time dependence determined solely by the characteristic function $\Gamma$. From Euler's formula $\cos(\Omega t) = \dfrac{e^{i\Omega t} + e^{-i\Omega t}}{2}$, the pitch angle diffusion coefficient can be written as
\begin{equation}
    D_{\mu \mu } = \frac{4\pi {\Omega}^{2}\left( 1-\mu ^{2}\right)}{B_{0}^{2}}\ \int _{0}^{\infty } dk_{\parallel } \ g( k_{\parallel}) \int _{0}^{\infty } dt\ \left( \frac{e^{i{\Omega}t} +e^{-i{\Omega}t}}{2} \right)  \Gamma( k_{\parallel } ,\ t) \label{eq:DmumuOmega}.
\end{equation}
Since we focus on the slab turbulence characteristic function, the subscript $\parallel$ is omitted for convenience in the following text. The resonance function describes the interaction between particles and turbulence, which can be identified from Eq \eqref{eq:DmumuOmega} as
\begin{equation}
    K_{\pm } =1/2\  Re\int _{0}^{\infty } e^{\pm i{\Omega}t} \ \Gamma 
     ( k  ,\ t) \ dt.
\end{equation}
The first term in the integrand is related to the particle's relativistic gyro frequency $\Omega = qB/(\gamma m)$, and thus is determined by the mean magnetic field, particle speed, charge and mass. The second term, the characteristic function $\Gamma$, depends on the turbulence wave number and guiding center position. It is a function of pitch angle and particle speed. 
Due to the dispersion of the pitch angle and the resulting parallel speed, 
the guiding center is perturbed about the mean position $\langle z \rangle$ as it moves along the magnetic field. 
In SOQLT, with the assumption that the guiding center parallel to the mean magnetic field direction follows a shifted Gaussian function, the characteristic function $\Gamma$ becomes
\begin{equation}
G(z)=\frac{1}{\sigma _{z}\sqrt{2\pi }} e^{\frac{-( z-\langle z\rangle )^{2}}{2\sigma _{z}^{2}}},
\end{equation}
where $\langle z\rangle$ and $\sigma_z^2 = \langle (z-\langle z \rangle)^2 \rangle$ are the mean and the variance of $z$ respectively. The characteristic function $ \Gamma$ \ can then be written as
\begin{equation}
    \Gamma(\vec{k} ,t) = \langle e^{ikz} \rangle =\int _{-\infty }^{\infty } dz\ G( z) \ e^{ik z} = e^{ik \langle z\rangle } \ e^{-\ \frac{ k^{2} \sigma _{z}^{2} }{2}}, \label{eq:Gamma}
\end{equation}
where $\Gamma(k ,t)$ depends on the ensemble averaged position of the particle $\langle z \rangle$
and the width of the Gaussian function $\sigma _{z}^{2}$. It should be noted that the characteristic function in QLT can be recovered by setting $\sigma_z=0$, so that
\begin{equation}
    \Gamma_{QLT}( \vec{k} ,\ t) =\ e^{ik \langle z\rangle }. \label{eq:Gammaqlt}
\end{equation}
Comparing Eq \eqref{eq:Gamma} and \eqref{eq:Gammaqlt}, we can see that if $k$ or $\sigma_z$ is large, the difference between the resonance function for SOQLT and QLT becomes large. From the view of QLT, lower energy particles as well as particles with large pitch angles will resonate with turbulence with large wave number. Furthermore, from the motion equation Eq \eqref{eq:dvdz}, a strong turbulent magnetic field gives rise to a large variance in $z$. We can expect that there will be considerable differences in the calculated diffusion coefficients. 

To proceed, unperturbed orbits are adopted to calculate $z$ and $\sigma_z^2$. Integrating the motion Eq \eqref{eq:dvdz} twice, we obtain
\begin{equation}
   z( t ) - v\mu t  = \frac{\Omega}{B_{0}}\int_{0}^{t } d\tau\int_{0}^{\tau } dt  [ v_{x}( t) \delta B_{y}( t) -\ v_{y}( t) \delta B_{x}( t)] , \label{eq:zt}
\end{equation}
and we adopt unperturbed $v_{x,y}$ in Eq \eqref{eq:vxvy} ignoring the presence of turbulence. 
By applying the ensemble average operator, $\langle \delta B_{x,y}\rangle=0$, the integrand becomes zero, and 
\begin{align}
    \langle z( t ) \rangle \ =\ v\mu t ,
\end{align}
as a consequence of the ensemble averaged force being equal to zero.
As $\sigma_z^2 = \langle (z-\langle z \rangle)^2 \rangle$, we square the right hand side of Eq \eqref{eq:zt} and apply the ensemble average
\begin{equation}
\sigma_{z}^{2}( t ) = \frac{\Omega^{2}  v^{2}  \left( 1-\mu ^{2}\right)}{B_{0}^{2}} \int_{0}^{t } d\tau_{1} \int_{0}^{t} d\tau_{2}\int_{0}^{\tau_{1}} dt_{1}\int_{0}^{\tau _{2}} dt_{2} \cos( \Omega( t_{1}  - t_{2}))  \langle \delta B_{x}( t_{1}) \delta B_{x}^{*}( t_{2}) \rangle.
\end{equation}
By adopting the characteristic function of QLT, $\Gamma_{QLT}(k_{\parallel}, t) = e^{ikv\mu t}$, the correlation function is given by from Equation \eqref{eq:corr_fun}
\begin{equation}
    \langle \delta B_{x}( t_{1}) \delta B_{x}^{*}( t_{2}) \rangle  = 4\pi  \int _{0}^{\infty } dk \ g( k) \ e^{ik v\mu ( t_{1} -t_{2})}.
\end{equation}
Thus,
\begin{equation}
    \sigma _{z}^{2}( t) = \frac{4\pi {\Omega}^{2}  v^{2}  \left( 1-\mu ^{2}\right)}{B_{0}^{2}}  \int _{0}^{\infty } dk\  g(k)\ (M_{+}( t)  + M_{-}( t)),
    \label{eq:gaussian_broadening}
\end{equation}
where $M_{\pm}$ is the time dependent resonance function for $\sigma _{z}^{2}( t)$,
\begin{align}
    M_{\pm }( t) &= \int_{0}^{t } d\tau_{1} \int_{0}^{t} d\tau_{2}\int_{0}^{\tau_{1}} dt_{1}\int_{0}^{\tau _{2}} dt_{2} \cos( \Omega( t_{1}  - t_{2}))e^{ikv\mu(t_1-t_2)} \\
    &= \frac{1-cos( \beta _{\pm } t)}{\beta _{\pm }^{4}} \ -\ \frac{sin( \beta _{\pm } t)}{\beta _{\pm }^{3}} \ t\ +\ \frac{t^{2}}{2\beta _{\pm }^{2}} ,
\end{align}
and $\beta _{\pm } =k_{\parallel } v\mu  \pm \Omega$.
In the case that $\mu =0$ and $t\gg 1/\Omega$, the last term dominates
\begin{equation}
   \lim_{t\gg 1/\Omega, \mu=0} M_{+} + M_- = 1/(\Omega t)^2,
\end{equation}
and the corresponding $\sigma_z^2$ is
\begin{equation}
    \lim_{t\gg 1/\Omega, \mu=0} \sigma_z^2 = \frac{v^2}{2} \frac{\delta B^2}{B_0^2}t^2. \label{eq:Omegat}
\end{equation} 
This is the so called $90^{\circ}$ large-time-approximation. 

With the approximation that magnetic momentum is conserved, \cite{Volk1975} derives the Root Mean Square (RMS) deviation of $v_{\parallel}$ from its unperturbed value, finding that
\begin{equation}
    \langle ([\Delta (v_{\parallel}^2) ]^2)^{1/4} \rangle = v_{\perp} \left( \frac{\langle (|\bm B|-B_0)^2\rangle}{B_0^2} \right)^{1/4},
\end{equation}
where $\bm B$ is the total magnetic field.
\cite{Volk1975} found that for the case of uncorrelated Alfv\'en and magnetosonic waves, the RMS deviation is dominated by the parallel components of the magnetosonic perturbation. 
Based on this work, \cite{Yan2008} developed a similar nonlinear diffusion theory where $\sigma_z^2=\langle \Delta (v_{\parallel}^2) \rangle t^2=v_{\perp}^2 (\langle \delta B_{\parallel}^2 \rangle/B_0^2)^{1/2}t^2$. They attribute $\delta B_{\parallel}$ to the slow modes (also known as the pseudo-Alfv\'en modes in the incompressible limit). In their model, the information relates time dependent phase for both particle speed and magnetic field is missed.   

Finally, we have all the ingredients to calculate the resonant function for SOQLT,
\begin{equation}
    K_{\pm } \ =1/2\ Re \int _{0}^{\infty } e^{i (\pm \Omega + kv\mu)t} e^{-k^2\sigma_z^2(t)/2} dt . \label{eq:broaden}
\end{equation}
As discussed before for Eq \eqref{eq:Dmumu}, the integration over time $t$ should be truncated to ensure that the assumption that $\mu$ is nearly constant be reasonable. Here, the exponential term $e^{-k^2\sigma^2_z/2}$ may play this role as it approaches zero rapidly with increasing $t$. Setting $\sigma_z=0$ for QLT, the resonant function becomes $\pi/2\ \delta (kv\mu \pm \Omega)$. At pitch angles close to $90^{\circ}$, the required resonant wave number approaches infinity. As there is no turbulence energy distribution with such high wave number ($P(k \to \infty)=0$), particles cannot reverse their direction. This issue is solved in the SOQLT by broadening the resonant function, and can be easy to see by substituting $\sigma_z^2$ with the $90^{\circ}$ large time approximation Eq \eqref{eq:Omegat} into Eq \eqref{eq:broaden},
\begin{align}
      \lim_{\mu \to 0}  K_{\pm }^{( 2)} &= 1/2\ Re \int _{0}^{\infty } e^{i (\pm \Omega + kv\mu)t} e^{-k^2\sigma_z^2(t)/2} dt = 1/2 \ Re \int _{0}^{\infty } e^{i (\pm \Omega + kv\mu)t} e^{-v^2k^2t^2\delta B^2/4 B_0^2}\ dt\\
        & =1/2 \frac{\sqrt \pi}{v \ k\ (\delta B/B_0)} \exp\left[-\left( \frac{\mu \pm \Omega/(kv)}{\delta B/B_0}\right)^2\right]. 
\end{align}
 In the QLT limit \(\delta B/B_0 \to 0\), the resonance function reduces to a Dirac delta, which enforces exact resonance and yields the well‑known “\(90^\circ\)” result \(D_{\mu\mu}(\mu=0)=0\). In the heliosphere, however, \(\delta B/B_0\) is finite, so the resonance function is broadened and turbulence at intermediate wavenumbers can scatter particles with \(\mu\simeq 0\). This behavior appears naturally in SOQLT, which predicts substantial \(D_{\mu\mu}\) near \(\mu=0\). Consistent with this, test‑particle simulations report strong scattering around \(\mu=0\) \citep[e.g.,][]{Qin2009,Shalchi2009}. Our SOQLT calculation agrees with those simulations, whereas QLT predicts \(D_{\mu\mu}(\mu=0)=0\), highlighting the limitation of the QLT \(\delta\)‑resonance in finite‑amplitude turbulence.

\clearpage

\bibliography{main_1}{}
\bibliographystyle{aasjournal}

\end{document}